\documentclass[preprint,onecolumn,aps,nofootinbib]{revtex4}
\usepackage{graphicx}
\pdfoutput=1
\usepackage[colorlinks=true,linkcolor=blue,urlcolor=blue,filecolor=black,citecolor=red,pdfstartview=FitV,
pdftitle={},pdfsubject={},pdfkeywords={},pdfpagemode=None,bookmarksopen=true]{hyperref}
\usepackage{extarrows}
\usepackage{epsfig}
\usepackage{amsmath}
\usepackage{amsfonts}
\usepackage{amssymb}
\usepackage{color}%
\usepackage{dcolumn}
\providecommand{\U}[1]{\protect\rule{.1in}{.1in}}

\newcommand{\f}{\begin{equation}}
\newcommand{\ff}{\end{equation}}
\newcommand{\fa}{\begin{eqnarray}}
\newcommand{\ffa}{\end{eqnarray}}

\allowdisplaybreaks[4]
\begin{document}
\title{Holographic superconductor with momentum relaxation and Weyl correction }
\author{Yi Ling $^{1,2,3}$}
\email{lingy@ihep.ac.cn}
\author{Xiangrong Zheng $^{1}$}
\email{xrzheng@ihep.ac.cn}
\affiliation{$^1$ Institute of High Energy Physics, Chinese Academy of Sciences, Beijing 100049, China\ \\
$^2$ Shanghai Key Laboratory of High Temperature Superconductors,
Shanghai, 200444, China\ \\
$^3$ School of Physics, University of Chinese Academy of Sciences,
Beijing 100049, China}
\begin{abstract}

We construct a holographic model with Weyl corrections in five
dimensional spacetime. In particular, we introduce a coupling term
between the axion fields and the Maxwell field such that the
momentum is relaxed even in the probe limit in this model. We
investigate the Drude behavior of the optical conductivity in low
frequency region. It is interesting to find that the incoherent
part of the conductivity is suppressed with the increase of the
axion parameter $k/T$, which is in contrast to other holographic
axionic models at finite density. Furthermore, we study the
superconductivity associated with the condensation of a complex
scalar field and evaluate the critical temperature for
condensation in both analytical and numerical manner. It turns out
that the critical temperature decreases with $\tilde{k}$,
indicating that the condensation becomes harder in the presence of
the axions, while it increases with Weyl parameter $\gamma$. We
also discuss the change of the gap in optical conductivity with
coupling parameters. Finally, we evaluate the charge density of
the superfluid in zero temperature limit, and find that it
exhibits a linear relation with
$\tilde{\sigma}_{DC}(\tilde{T_c})\tilde{T_c}$, such that a
modified version of Homes' law is testified.
\end{abstract}
\maketitle

\section{Introduction}\label{section1}
The high $T_c$ superconductivity in some novel materials such as
Cuprates and heavy Fermion compounds involves the strong couplings
of electrons, thus beyond the traditional BCS theory which
describes the electron-phonon interaction very well. Recent
progress in the AdS/CMT duality has provided powerful tools for
understanding novel phenomena in strongly coupled system
\cite{Hartnoll:2008vx,McGreevy:2009xe,Herzog:2009xv,Iqbal:2011ae,Hartnoll:2008kx,Gubser:2008px,Horowitz:2010gk,Hartnoll:2009sz}.
Especially, a holographic description of the condensation was
originally proposed in \cite{Gubser:2008px} by Gubser and a
holographic model for the superconductor was firstly constructed
in \cite{Hartnoll:2008vx,Hartnoll:2008kx} by Hartnoll, Herzog and
Horowitz. One key ingredient in this holographic model is
introducing the spontaneous breaking of $U(1)$ gauge symmetry in
the bulk geometry, which is analogous to the mechanism of s-wave
superconductors. We refer to Refs.
\cite{Horowitz:2010gk,Hartnoll:2009sz,Cai:2015cya} for a
comprehensive review on holographic superconductors.

It is very interesting to enrich the holographic setup to
investigate various novel phenomena observed in condensed matter
experiments. For instance, a Weyl term composed of the couplings
of the Weyl tensor and the Maxwell field was introduced in
\cite{Ritz:2008kh,Myers:2010pk} and then has extensively been
investigated in holographic literature
\cite{Sachdev:2011wg,Wu:2010vr,Momeni:2011ca,WitczakKrempa:2012gn,WitczakKrempa:2013ht,Witczak-Krempa:2013aea,
Witczak-Krempa:2013nua,Katz:2014rla,Ling:2016dck,Wu:2016jjd}.
Since the self duality of the Maxwell field is violated due to the
presence of the Weyl term, it provides a novel mechanism for
understanding the condensed matter phenomena with the breakdown of
the electromagnetic self-duality from a holographic perspective.
Historically, this model also provides a way of turning on a
nontrivial frequency dependence for the optical conductivity of
the dual QFT, which has been studied with small parameter
perturbation theory as well as the variational method in
\cite{Myers:2010pk,Sachdev:2011wg,Wu:2010vr,Momeni:2011ca,WitczakKrempa:2012gn}.
However, to obtain a finite DC conductivity over a charged black
bole background, one essential step is to break the translational
invariance such that the momentum is not conserved. Until now
there are many ways to introduce the momentum relaxation or
momentum dissipation in holographic approach. It can be
implemented by spatially periodic sources
\cite{Horowitz:2012ky,Horowitz:2012gs,Horowitz:2013jaa,Ling:2013aya,Ling:2013nxa,Ling:2014saa},
helical and Q-lattices
\cite{Donos:2012js,Donos:2014oha,Donos:2013eha,Donos:2014uba,Ling:2014laa},
spatially linear dependent axions
\cite{Andrade:2013gsa,Gouteraux:2014hca,Kim:2014bza,Cheng:2014qia,Ge:2014aza,Mozaffara:2016iwm},
or massive
gravitons\cite{deRham:2010kj,Hassan:2011hr,Vegh:2013sk,Blake:2013bqa,Blake:2013owa,Davison:2013jba}(for
a brief review we refer to Ref.\cite{Ling:2015ghh}). In the
presence of slow momentum relaxation, it is found that the optical
conductivity exhibits a Drude behavior in the low frequency
region, which has become a widespread phenomenon in holographic
models. However, when the momentum relaxation becomes strong, the
discrepancy from the Drude formula can be observed even in the low
frequency region, leading to the incoherence of the conductivity.
Currently, it is still a crucial issue to understand the
coherent/incohenrent behavior of metal in both condensed matter
physics and holographic gravity
\cite{Kim:2014bza,Ge:2014aza,Hartnoll:2014lpa,Davison:2014lua,Davison:2015bea,Davison:2015taa,Zhou:2015qui}.

In this paper we intend to construct a holographic model with
momentum relaxation in the presence of the Weyl term. Moreover,
inspired by recent work in \cite{Gouteraux:2016wxj}, we intend to
provide a novel scheme to introduce the momentum relaxation by
considering the coupling between the axions and the Maxwell field.
In this way we could consider the coherence/incoherence of the
metal even in the probe limit where the charge density is
vanishing in the background. In contrast to usual
holographic models with axions where the incoherence of the
conductivity becomes manifest when the momentum relaxation becomes
strong, we find in our model the portion of incoherent
conductivity is suppressed with the increase of the axion
parameter $k/T$.

In the second part of this paper we will construct a holographic
model for s-wave superconductor by the spontaneous breaking
of $U(1)$ gauge symmetry. We will investigate the condensation of
the complex scalar field and evaluate the critical temperature as
well as the energy gap which may vary with the strength of the
momentum relaxation. Another motivation of our current work is to
testify the Homes' law in the holographic perspective. In
condensed matter literature \cite{Homes2004,Homes2005}, it has
been shown by experiments that for a large class of
superconductivity materials there exists an elegant empirical law
linking the charge density of superfluid at zero temperature to
the $DC$ conductivity near the critical temperature, which now is
dubbed as Homes' law. This law discloses that regardless of the
structure of materials, the superconductivity always exhibits a
universal behavior as
\begin{equation}\label{homeslaw}
  \rho_s (T=0) = C \sigma_{DC}(T_c) T_c,
\end{equation}
where the constant $C$ is found to be about $4.4$ for in-plane
high-$T_c$ superconductors and clean BCS superconductors, while
for $c$-axis high-$T_c$ materials and BCS superconductors in dirty
limit $C=8.1$. For organic superconductors the $C=4\pm 2.1$
\cite{Homes2013}. On the theoretical side, a preliminary
understanding on the Homes' law of high-$T_c$ superconductors was
proposed with the use of the notion of Planckian dissipator in
\cite{Zaanen2004}. However, for the conventional superconductors
which are subject to the Homes' law as well, a corresponding
interpretation in theory is still missing. An alternative
mechanism with double timescales proposed to understand Homes' law
can also be found in \cite{Phillips2005}. Nevertheless, people
believe that the present stage is still far from a complete
understanding on the Homes' law in theory. In particular, for high
temperature superconductivity, it is believed that the BCS theory
breaks down and some novel techniques should be developed to treat
the many-body system which is strongly coupled. Holography has
rendered us a powerful tool to address such an open problem.
Therefore, recently it has been becoming very intriguing to
testify if the Homes' law would be observed in holographic
approach. Earlier attempts in this route can be found in
\cite{Erdmenger:2012ik,Erdmenger:2015qqa,Kim:2016hzi,Kim:2016jjk}.
In this paper we intend to demonstrate that a modified Homes' law
can be observed in our model as well.

This paper is organized as follows. The holographic setup of our
model is given in section \ref{section2}. Then we investigate the
electric transport properties of the dual field theory in section
\ref{section3}, focusing on the coherent-incoherent transition
with the strength of axions. In section \ref{section4} we turn to
the superconductivity of this model. The critical temperature for
condensation and the energy gap are evaluated. Particularly, we
will focus on the test of Homes' law based on the transport
properties of the superconductivity. Some open questions and
possible development are discussed in section \ref{section5}. As a
byproduct, we present an efficient way to get rid of the effects
due to the presence of nonanalytic terms in numerical simulation
in the Appendix.

\section{Holographic Setup}\label{section2}

The holographic superconductors with momentum relaxation and
dissipation have been investigated in various models, such as
\cite{Horowitz:2013jaa,Ling:2014laa,Andrade:2014xca,Erdmenger:2015qqa,Kim:2016hzi,Kim:2016jjk,Kim:2015dna,Baggioli:2015zoa}
. Here we consider a holographic model in Einstein-Maxwell-Axion
theory with a Weyl correction in five dimensional spacetime. The
total Lagrangian is given by
\cite{Gouteraux:2016wxj,Baggioli:2016oqk}
  \begin{equation}\label{1}
    \begin{aligned}
  \mathcal{L} = R+\frac{12}{L^2}-\frac{1}{4}\left(1+\mathcal{K}Tr[X]\right)F^2-{\left|\nabla\psi-ieA\psi\right|}^2-m^2{\left|\psi\right|}^2
  +{\gamma}L^2C_{\mu\nu\rho\sigma}F^{\mu\nu}F^{\rho\sigma},
   \end{aligned}
  \end{equation}
where $X^{\mu}
_{\nu}=\frac{1}{3}g^{\mu\tau}\partial_{\tau}{X^I}\partial_{\nu}{X^I}$
with ${X^I}$ being axion fields, which constitute massive term of
graviton. $C_{\mu\nu\rho\sigma}$ is the Weyl tensor which is
coupled to the Maxwell field with a coupling parameter $\gamma$.
In this action we have also introduced a coupling term between the
axions and Maxwell fields as proposed in \cite{Gouteraux:2016wxj}.
Following the analysis in \cite{Gouteraux:2016wxj}, we will set
the coupling constant $\mathcal{K}=\frac{1}{7}$ throughout this
paper. In addition, from \cite{Ritz:2008kh} we know the
value of $\gamma$ has to obey the following constraint for the
sake of causality and stability of the system
\begin{equation}\label{a5}
\begin{aligned}
-\frac{L^2}{16}<\gamma <\frac{L^2}{24}.
 \end{aligned}
 \end{equation}
Furthermore, for simplicity we set $m^2=-\frac{3}{L^2}$ and $L=1$.

In the presence of the Weyl correction, usually it is very hard to
obtain the analytical solutions with backreaction, although the
approximate solution could be obtained when the Weyl parameter is
very small\cite{Ling:2016dck}. In current paper we will only
consider the probe limit of the system. That is to say, the
gravity is decoupled from the matter fields. We fix the background
as an AdS-Schwarzschild black brane whose metric reads as

\begin{equation}\label{b43}
\begin{aligned}
ds^2
=\frac{{r_H}^2}{z^2}\left(-f(z)dt^2+dx^2+dy^2+dw^2\right)+\frac{dz^2}{z^2f(z)},
\end{aligned}
\end{equation}
where $f(z)=1-z^4$ with $0<z<1$. $r_H$ is the position of
the black hole horizon and we will set it as unit in numerics
throughout this paper. The Hawking temperature of the black hole
is given as $T=\frac{r_H}{\pi}$.

\section{Conductivity in normal phase}\label{section3}

In this section we will investigate the electrical transport
behavior of the dual field theory without condensation, namely
$\psi=0$. Firstly, it is easy to see that
$X^I=k\delta^I_ix^i$, with $i$ running over boundary space
indices, is a solution to the equations of motion for axion
fields. We will consider the conductivity of the gauge field in
one spatial direction, say x-direction, over such a background
with momentum relaxation. In the absence of the condensation, the
$x$-component of the gauge field is decoupled from other fields
except axions $X^I$ and thus we can only turn on the electric
field in this direction. Consider a perturbation with a form
$A_x(z)e^{i\omega t}$ in linear response theory, we have a single
perturbation equation as
\begin{equation}\label{a33}
\begin{aligned}
A_x''(z)+\left(\frac{3}{z}+\frac{g'(z)}{g(z)}+\frac{2 \left(\gamma
z^5 g^{(3)}(z)+6 \gamma  z^4 g''(z)+6 \gamma  z^3
g'(z)+3\right)}{z h(z)}\right)A_x'(z)\\+\frac{\omega ^2}{z^4
g(z)^2}A_x(z)=0
\end{aligned}
\end{equation}
where $h(z)=2 \gamma z^4 g''(z)+8 \gamma  z^3 g'(z)+4 \gamma  z^2
g(z)-3 \mathcal{K} k^2 z^2-3$. Notice that in five dimensional
space time,  ${A_x}(z)$ has the following asymptotical behavior at
spatial infinity $z=0$
\begin{equation}\label{a37}
\begin{aligned}
A_x(z)=A_x^0+A_x^2{z^2}-\frac{\left(\omega ^2 A_x^0\right)\log
\left(\frac{z}{\Lambda}\right) { z^2}}{2},
\end{aligned}
\end{equation}
where $\Lambda$ is an arbitrary constant. The logarithmic
divergency can be removed by introducing a counterterm, which is
\cite{Taylor:2000xw,Horowitz:2012gs}
\begin{equation}\label{a38}
\begin{aligned}
S_{ct}=\frac{L\log (\nu  z)}{4}\int \sqrt{-\gamma } F^2 \,
dx^4.
\end{aligned}
\end{equation}
Then following the standard holographic dictionary, we may obtain
the optical conductivity for the dual field theory as
\begin{equation}\label{a40}
\begin{aligned}
\sigma (\omega )=\frac{1}{i L^3  \omega } \left( \frac{2
A_x^2}{A_x^0}+ \omega ^2 \left(\log \left(\frac{\Lambda }{\nu
}\right)-\frac{1}{2}\right)\right).
\end{aligned}
\end{equation}

Usually, the non-analytic behavior caused by the logarithmic term
brings some difficulties to guarantee the convergency in numerical
calculation and it becomes harder for one to extract the data for
the coefficients in the expansion (\ref{a37}) with enough
accuracy. Alternatively, we find that it is more numerically
efficient to replace $A_x(z)$ by $(1-z)^{\frac{-i\omega}{4}}
\left(a_x(z)-\frac{1}{2} \omega ^2 A_x^0 \frac{\log
(z)z^2}{2}\right)$ in perturbation equation (\ref{a33}), then the
equation for $a_x(z)$ is analytic everywhere and the conductivity
can be reexpressed in terms of $a_x(z)$ as

\begin{equation}\label{a41}
\begin{aligned}
\sigma (\omega )=\frac{1}{i L^3  \omega }\left(\frac{a_x^{''}(z\to
0 )}{a_x(z\to 0 )}+ \omega ^2 \left(\log \left(\frac{\Lambda }{\nu
}\right)-\frac{1}{2}\right)+\frac{\omega ^2}{16}+\frac{i \omega }{4}\right).
\end{aligned}
\end{equation}
The derivation from (\ref{a40}) to (\ref{a41}) is presented
in Appendix.
\begin{figure}[!h]
  \centering
  \includegraphics[height=4.5cm]{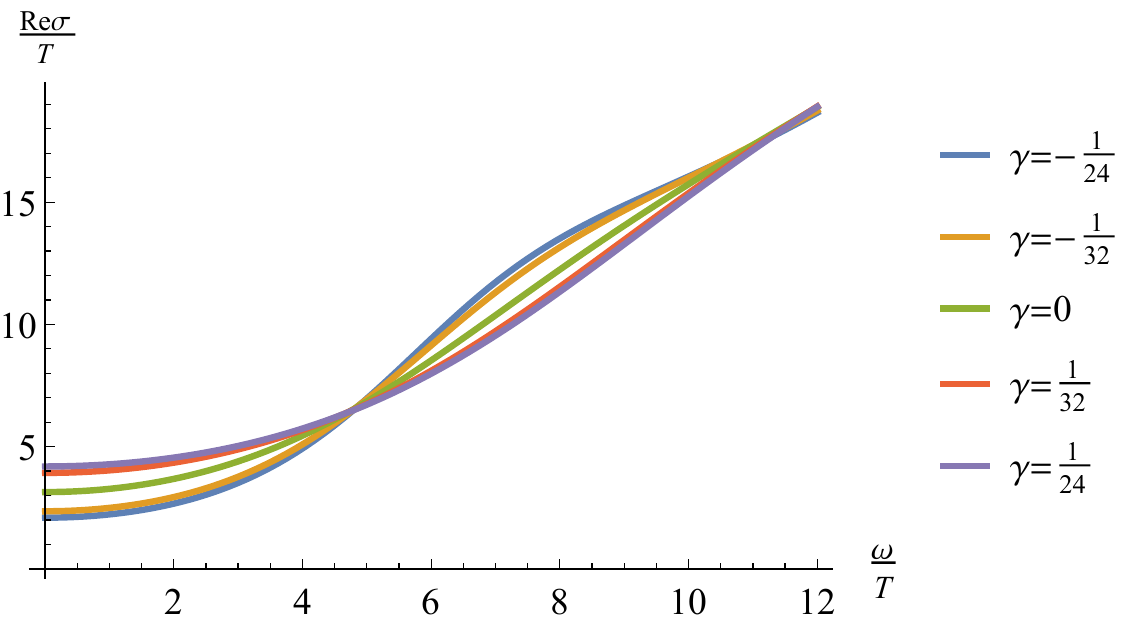}\qquad
  \includegraphics[height=4.5cm]{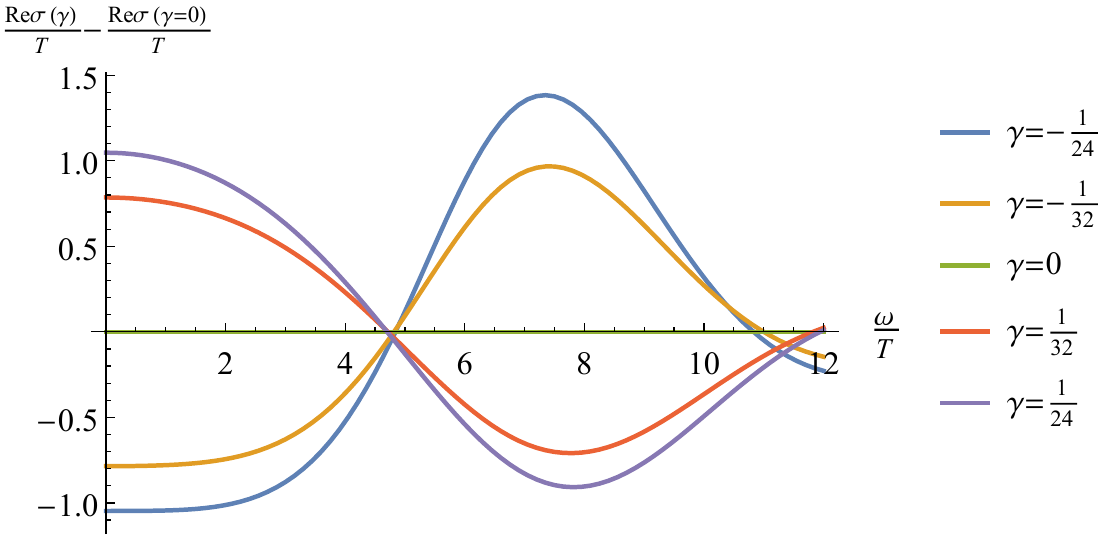}
  \caption{The real part of the conductivity for different values of $\gamma$ with $k=0$. }\label{RC}
\end{figure}

   \begin{figure}[!h]
  \centering
  \includegraphics[height=4.5cm]{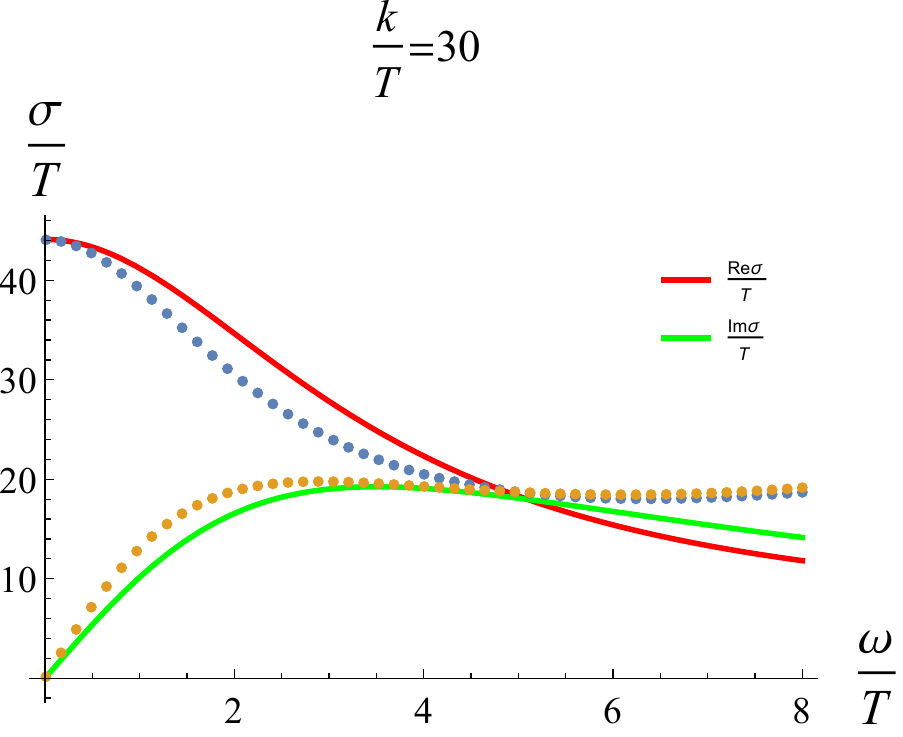}\qquad
  \includegraphics[height=4.5cm]{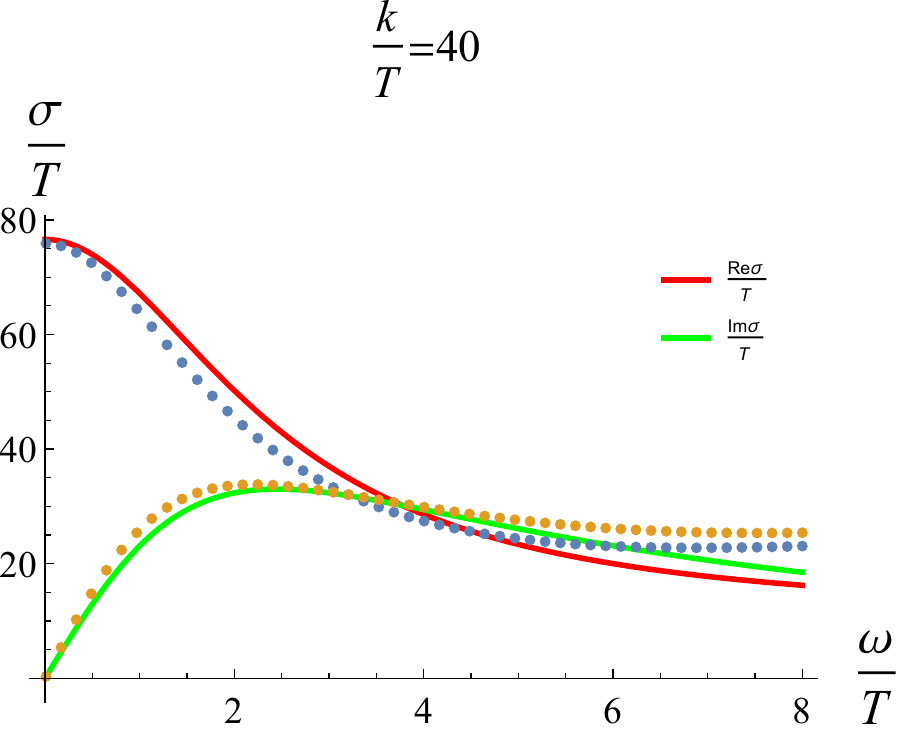}\qquad
  \includegraphics[height=4.5cm]{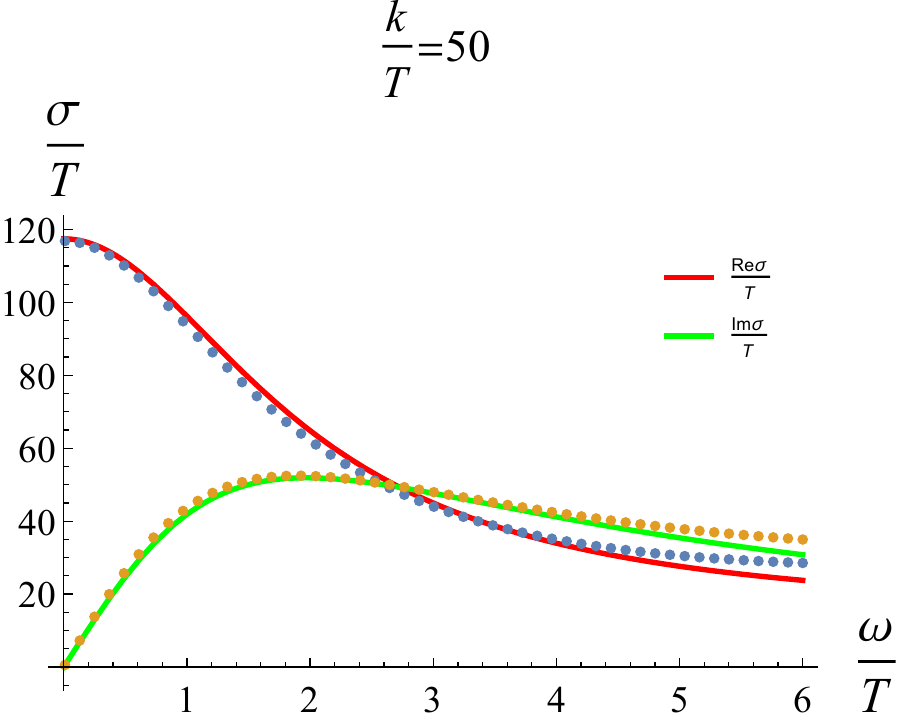}\qquad
   \includegraphics[height=4.5cm]{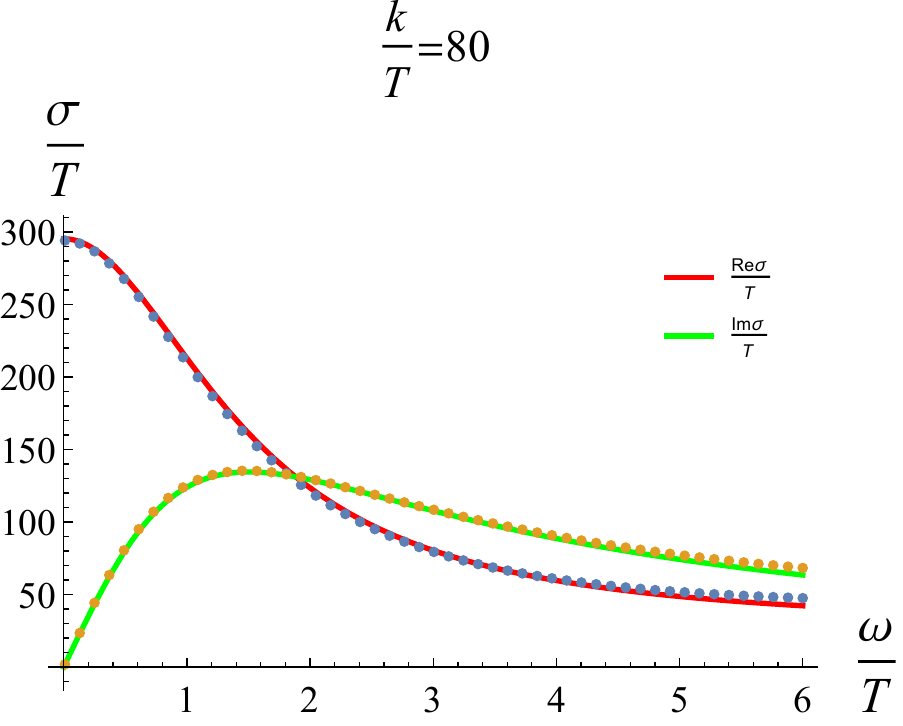}
  \caption{\label{Incoherent}
 The frequency dependence of conductivity for $k/T=30, 40, 50,$ and $80$ respectively
 with
$\gamma=0$. Dotted lines are numerical data, while solid lines are
fit with the modified Drude formula
$\sigma(\omega)=\frac{K\tau}{1-i \omega\tau}+\sigma_Q$.}
\end{figure}

Next we intend to demonstrate our numerical results for the
optical conductivity based on Eq.(\ref{a41}) in the case of
$\log\left(\frac{\Lambda }{\nu }\right)=0$.

\begin{itemize}
\item $k=0,\gamma\neq 0$. It is very interesting to notice that
both the axion parameter $k$ and Weyl parameter $\gamma$ appear
    in the same term in the perturbation equation, as shown in
    (\ref{a33}), implying that the momentum relaxation and the Weyl correction play a similar role in
    influencing the optical conductivity. Firstly, we illustrate the
    frequency behavior of the real part of the conductivity for various values of $\gamma$ with $k=0$, as plotted in Figure.\ref{RC}. It is nothing but the extension of the results in
    four dimensions presented in \cite{Myers:2010pk} to five dimensions. For zero
    frequency region, we observe the similar phenomena as
    disclosed in \cite{Myers:2010pk}. Namely, the DC conductivity increases with
    $\gamma$. Quantitatively, our numerical result of $\sigma_{DC}$
   agrees with the analytical result which is $\sigma_{DC}/T=(1+8\gamma)/T=\pi(1+8\gamma)$ \cite{Ritz:2008kh}. While for large frequency region, the conductivity
    becomes independent of the
    value of $\gamma$, and always has a linear relation with the frequency rather than a
    constant as in four-dimensional case. This is a reflection of the different dimensions of the boundary geometry\cite{Myers:2010pk,Wu:2016jjd,WitczakKrempa:2012gn}.
   Furthermore, if one subtracts the conductivity with $\gamma=0$,
   a peak (for $\gamma>0$) or a dip (for $\gamma<0$) can obviously
   be observed near the zero frequency region, as illustrated in the
   bottom plot of Fig.\ref{RC}, which is similar to the
   phenomenon in four dimensions. Interestingly, for small $\gamma$ region, it seems there
   exists a mirror symmetry around the horizontal axis if one changes $\gamma$ to $-\gamma$, which has
   previously been discussed in \cite{Myers:2010pk,WitczakKrempa:2012gn,WitczakKrempa:2013ht,Witczak-Krempa:2013aea,
Witczak-Krempa:2013nua,Ling:2016dck}. Furthermore, since the value
of $\gamma$ is
    subject to the constraint in (\ref{a5}), we find the Weyl term has
    limited impact on the conductivity. Specially, a Drude
    behavior in low frequency region can not be observed. However, if one ignores the restriction in (\ref{a5}),
    a Drude peak would emerge when $\gamma\gg1$, as discussed in \cite{Wu:2016jjd}. This is not surprising as we have pointed out that
    the axion parameter $k$ and Weyl parameter $\gamma$ do play similar roles in the perturbation
    equation. Finally, we remark that if one introduces
    some other Weyl terms with higher derivatives, then the additional coefficients involved may not be bounded
    as $\gamma$ such that a substantial Drude-like peak at small
    frequencies can be achieved as well by adjusting these coefficients, as discussed in
    \cite{Witczak-Krempa:2013aea}.

 \item $k\neq 0,\gamma= 0$. In this case the Weyl curvature term is vanished
 while the
 momentum is relaxed. Firstly, for
$\omega=0$, the real part of numerical conductivity is increasing
with $k/T$, which is consistent with the analytical result in
\cite{Gouteraux:2016wxj}. Secondly, we find a prominent Drude-like
peak in low frequency region of the optical conductivity, as shown
in FIG.\ref{Incoherent}. Numerically, we also find that when $k/T$
becomes large, the $\sigma_{DC}$ is increasing and the numerical
data can be fit well with a modified Drude formula, namely,
$\sigma(\omega)=\frac{K\tau}{1-i\omega\tau}+\sigma_Q$, where
$\sigma_Q$ is a real constant signalizing the incoherent
contribution to the conductivity. To quantitatively measure the
incoherent contribution to the total conductivity, we intend to
define a quantity $\sigma_Q/(K\tau)$ and plot its variation with
the momentum relaxation $k/T$, which is shown in the left plot of
Fig.\ref{figurea1}. Interestingly enough, we find the portion of
incoherent contribution is decreasing with the increase of $k/T$.
It indicates that the metallic phase of the dual system looks more
coherent in large $k/T$ region indeed\footnote{Strictly speaking,
we need consider their contributions to conductivity within a
frequency region by considering the ratio
$\int^{\omega_c}_0\sigma_Qd\omega/\int^{\omega_c}_0(\sigma_Q+K\tau/(1-i\omega\tau))d\omega$.
But when we choose the cutoff $\omega_c$ as the same order as
$1/\tau$, we find the same behavior can be observed as for the
quantity $\sigma_Q/(K\tau)$.}. Such behavior is in contrast to the
phenomenon observed in most previous holographic models with
axions, where a Drude-like peak can be observed even with
small $k/T$ and the incoherence of the metal becomes evident in
large $k/T$ region \cite{Kim:2014bza,Wu:2016jjd,Ge:2014aza}.
First of all. In usual lattice models or axion models, the
backreaction of lattice to the background is taken into account
such that the translational invariance is broken already prior to
the linear perturbations. Technically, the chemical potential
$\mu$ contributes a term $\mu^2/k^2$ in the usual expression for the
DC conductivity, leading to a prominent Drude peak even if the
lattice effect is weak (with tiny $k/T$). However, in our paper,
only the neutral background is taken into account under the probe
limit such that the effect of this term is absent. More
importantly, the difference results from the different coupling
manner of axion and Maxwell field. In our model the axion
fields do not contribute any independent terms such as the
kinematic term or potential term in the Lagrangian (\ref{1}). It
only appears as a term coupled to Maxwell field such that it
induces the momentum relaxation only at the linear response level.
In hence, the Drude behavior would not become manifest until the
momentum relaxation becomes strong with the axion parameter
$k/T$. Moreover, the coupling term plays a double
role in influencing the transport behavior. One is to control the
generation of electron-hole pairs which roughly speaking is
responsible for the incoherent part of conductivity,
reflected by the quantity $\sigma_Q$. The other one is to
induce the momentum relaxation, leading to coherent
conductivity, reflected by the quantity $K\tau$
\cite{Gouteraux:2016wxj}. With the increase of $k/T$ both
quantities $\sigma_Q$ and $K\tau$ increase, while as a result of
competition, the ratio $\sigma_Q/(K\tau)$ becomes smaller with
larger $k/T$. As a matter of fact, the enhancement of the
coherence can also be perceived if one evaluates the relaxation
time for different $k/T$. Our results indicate that it increases
with $k/T$, as shown in the left plot of FIG.\ref{figure10}.

\begin{figure}[!h]
  \centering
  \includegraphics[height=4.5cm]{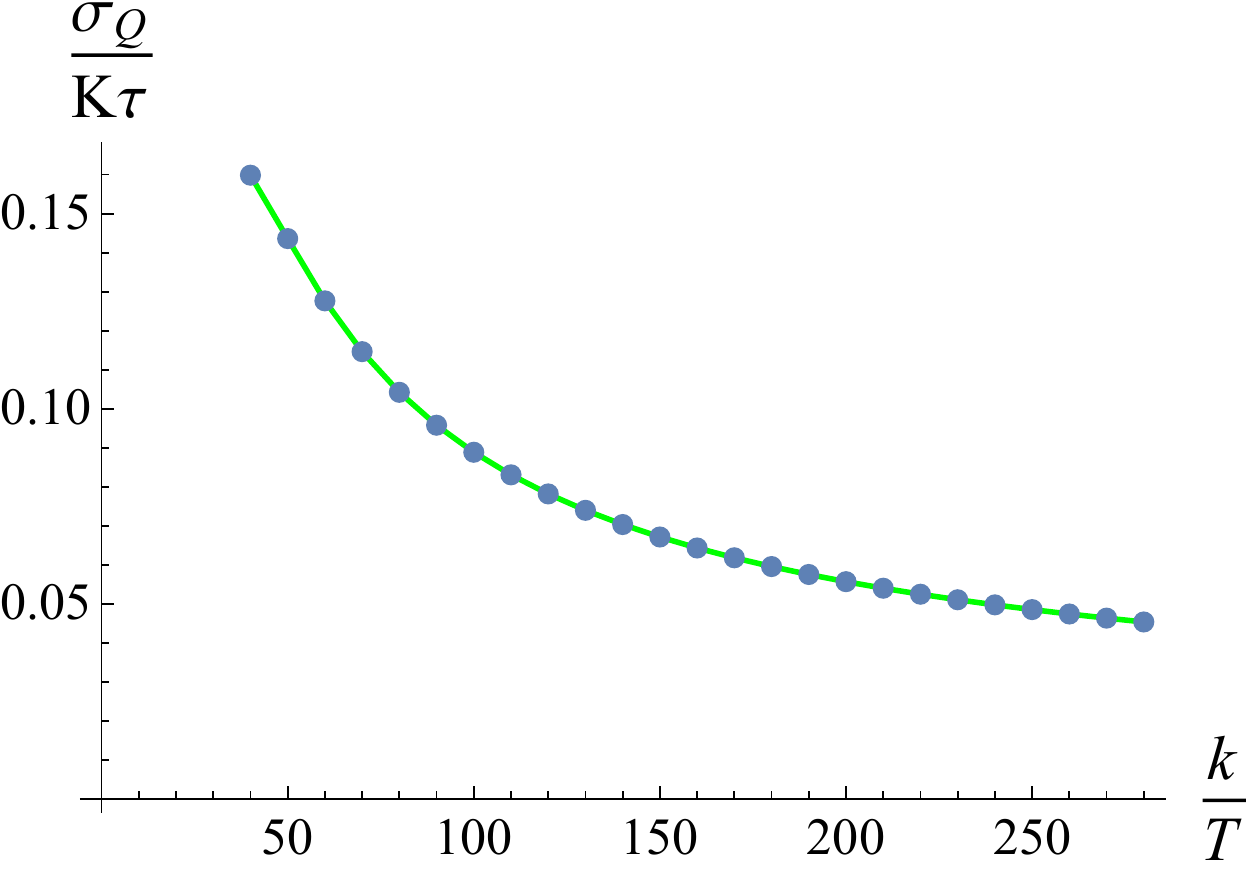}\qquad
    \includegraphics[height=4.5cm]{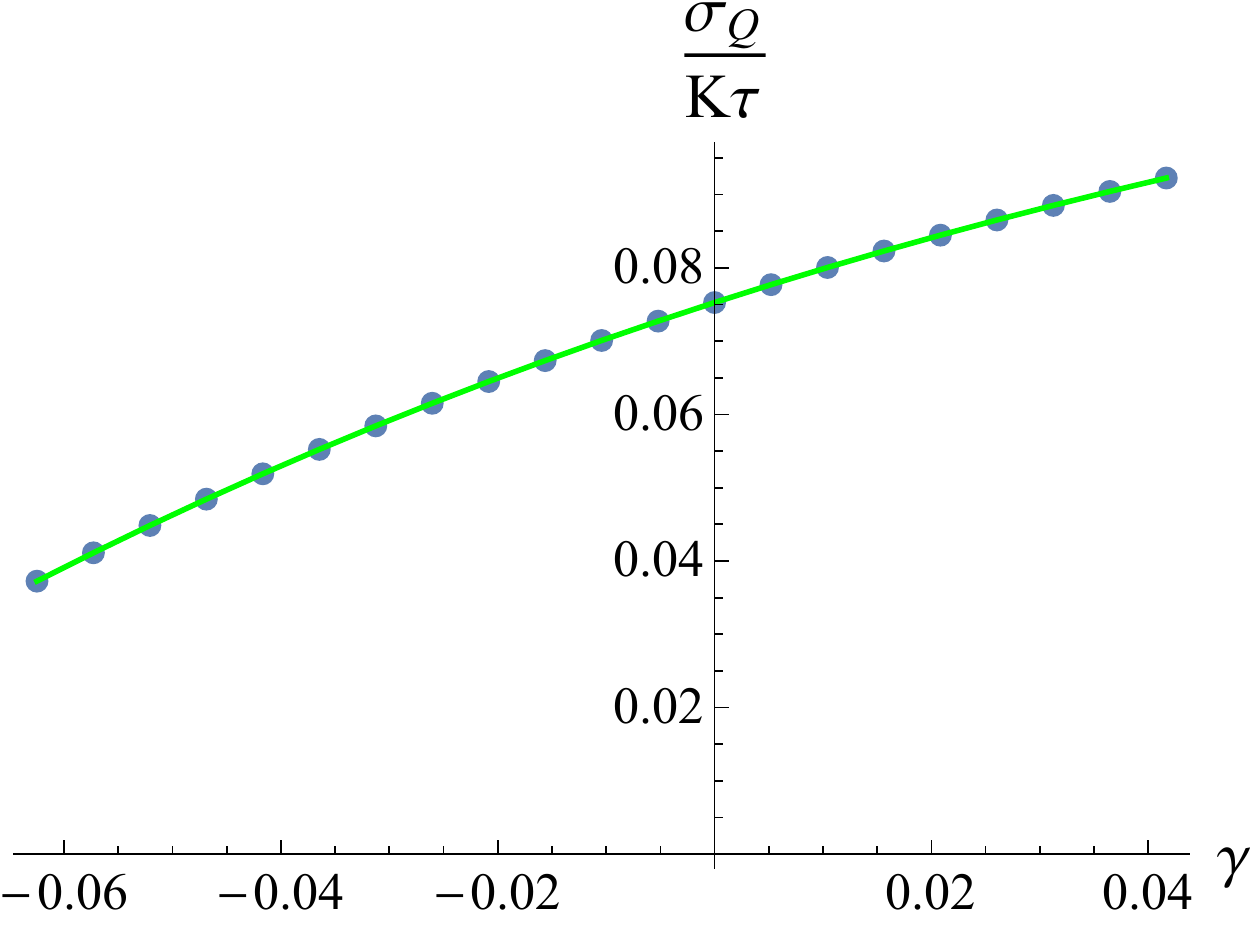}
  \caption{\label{figurea1}
The ratio of incoherent part to the coherent part in $DC$
conductivity. The left plot is for varying $k/T$ with $\gamma=0$,
while the right plot is for varying $\gamma$ with $k/T=25$}.
\end{figure}

\item $k\neq 0,\gamma\neq 0$. In this case the translational
invariance is broken. The optical conductivity has a similar
behavior as in above case when changing $k/T$ with $\gamma$ fixed,
since the Weyl parameter is constrained to take values in a small
region. On the other hand, if we change $\gamma$ with $k/T$ fixed,
we find that the ratio of the incoherent conductivity to the
coherent part increases with $\gamma$, as illustrated in the right
plot of FIG.\ref{figurea1}. In addition, our data indicate
that the relaxation time is linear with $\gamma$, and our fitted
result is $\tau T=0.463 \gamma +0.238$, which is shown in the
right plot of FIG.\ref{figure10}.

 \begin{figure}[!h]
  \centering
  \includegraphics[height=4.5cm]{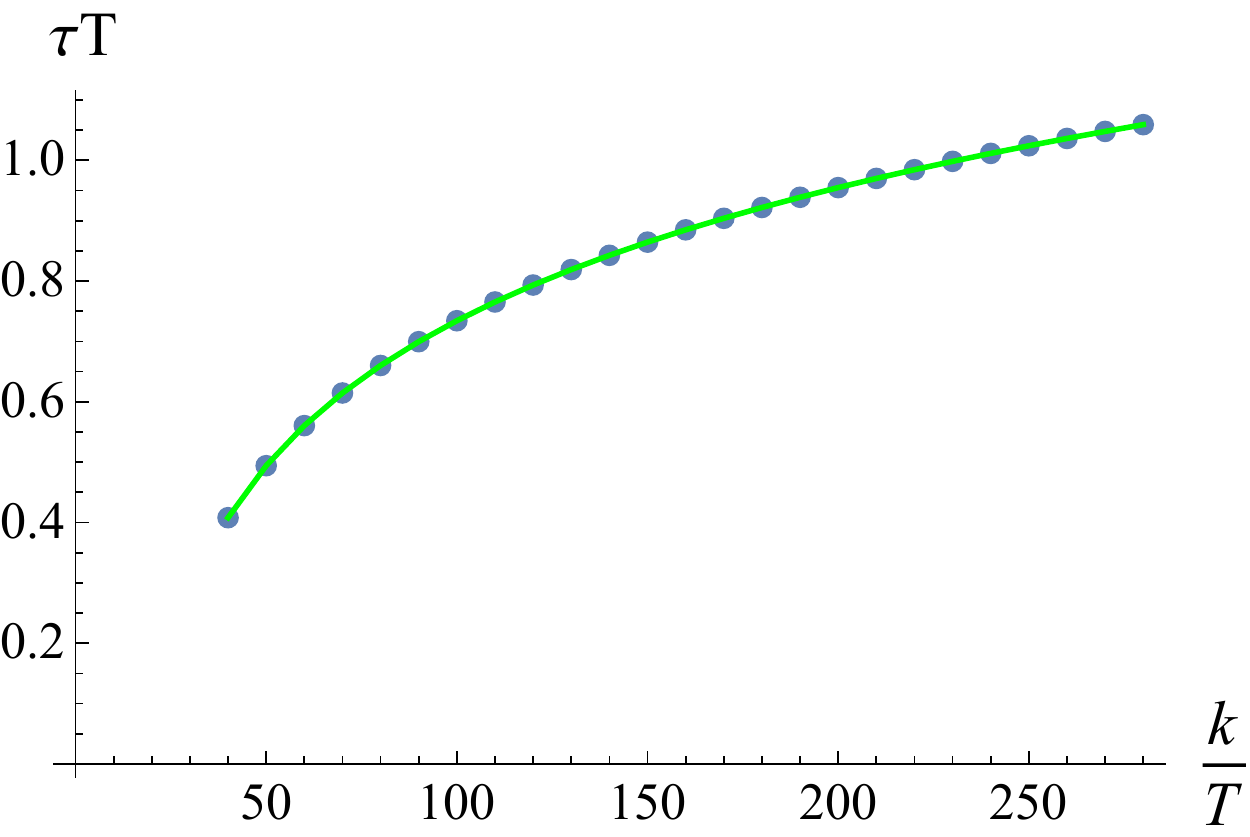}\qquad
   \includegraphics[height=4.5cm]{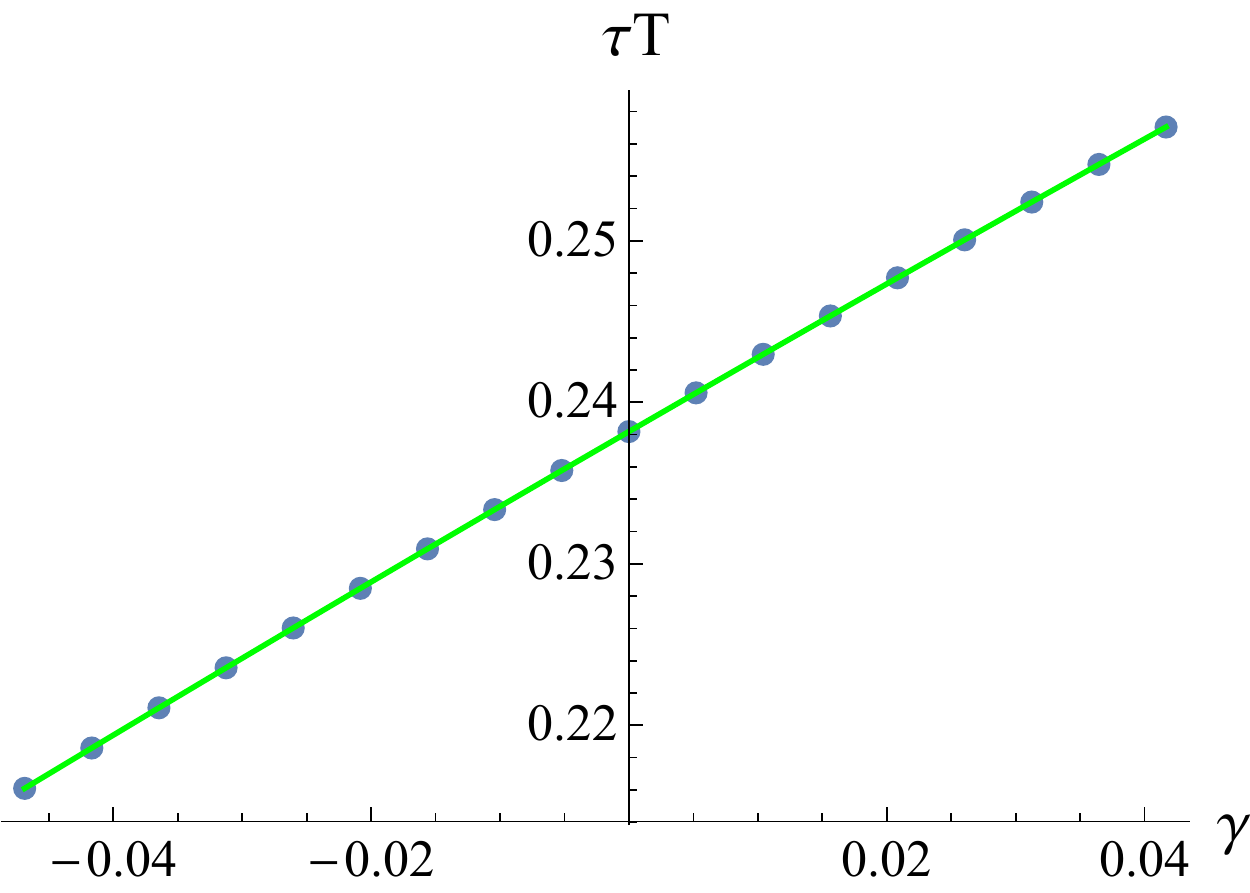}
  \caption{\label{figure10}
The relation between the relaxation time $\tau T$ and the
system parameters $k/T$ and $\gamma$. The left plot is for varying
$k/T$ with $\gamma=0$, while the right plot is for varying $\gamma$
with $k/T=25$.}
\end{figure}

\end{itemize}

\section{Conductivity in superconducting phase}\label{section4}

In the remainder of this paper we turn to investigate the
transport properties of the superconducting phase in this model.
We will turn on the complex scalar field and introduce the
spontaneous breaking of $U(1)$ gauge symmetry. In probe
limit, the system becomes simple and we intend to firstly
evaluate the critical temperature for the condensate in an
analytical way, and then compute the frequency behavior of
the conductivity with numerical analysis, focusing on the
evaluation of the energy gap and the test of Homes' law.
\subsection{Analytical part}\label{parta1}
In this subsection we will evaluate the critical temperature for
condensation closely following the procedures of matching method
as given in literature \cite{Gregory:2009fj,Kuang:2010jc,Roychowdhury:2012hp}.
The basic idea is
 to find an approximate solution to the equations of condensate by virtue of series expansion. We expand all the variables near the boundary $z=0$
 and the horizon $z=1$ separately which are subject to the equations of motion, then require these series expansion of the same variable is matched at
some intermediate location, for instance $z=\frac{1}{2}$.
\subsubsection{Expansion near the boundary and horizon}
After turning on the complex scalar field, the Lagrangian of the
matter part is given by
  \begin{equation}\label{e1}
    \begin{aligned}
  {\mathcal{L}}_2 =-\frac{1}{4}\left(1+\mathcal{K}Tr[X]\right)F^2-{\left|\nabla\psi-ieA\psi\right|}^2-m^2{\left|\psi\right|}^2
  +{\gamma}L^2C_{\mu\nu\rho\sigma}F^{\mu\nu}F^{\rho\sigma}.
   \end{aligned}
  \end{equation}
With the ansatz $A_\mu=(a(z),0,0,0,0)$, we derive the equations of
motion as follows

\begin{equation}\label{11}
\begin{aligned}
a''(z)+\left(\frac{2 \left(\gamma  z^5 g^{(3)}(z)+6 \gamma  z^4 g''(z)+6 \gamma  z^3 g'(z)-{r_H}^2\right)}{z e(z)}+\frac{1}{z}\right)a'(z)-\frac{2{r_H}^4 \psi (z)^2}{z^4 e(z) g(z)}a(z)=0,
\end{aligned}
\end{equation}

 \begin{equation}\label{10}
\begin{aligned}
\psi ''(z)+\frac{\psi '(z) \left(z^4 g(z) g'(z)- z^3 g(z)^2\right)}{ z^4 g(z)^2}+\frac{\psi (z) \left(a(z)^2+3 g(z)\right)}{ z^4 g(z)^2}=0,
\end{aligned}
\end{equation}

\begin{equation}\label{c9}
\begin{aligned}
\nabla _{\mu }\left[\left(\nabla ^{\mu }X^I\right) F^2\right]=0,
\end{aligned}
\end{equation}
where $g(z)=\frac{{r_H}^2}{z^2}\left(1-z^4\right)$, $e(z)=2 \gamma
z^4 g''(z)+8 \gamma  z^3 g'(z)+4 \gamma  z^2 g(z)+\mathcal{K} k^2
z^2+{r_H}^2$. It is easy to figure out that $X^I=k\delta^I_ix^i$
is a solution of the last equation of (\ref{c9}), where $i$ runs
over boundary space indices. With the use of EOM, we find the
asymptotical behavior of fields at $z=0$, which are
 \begin{equation}\label{15}
 \begin{aligned}
a(z)=\mu-qz^2+\cdots \qquad \qquad         \psi(z)={\psi_-}z^{\lambda_-}+\cdots+{\psi_+}z^{\lambda_+}+\cdots,
\end{aligned}
\end{equation}
where $\lambda_+=3$, $\lambda_-=1$.  On the other hand, with the
use of EOM we find $a(1)=0$ on the horizon. We require that the
Maxwell field and scalar field are regular at horizon, such that
we expand $a(z)$ and $\psi(z)$ near horizon as follows
 \begin{equation}\label{16}
 \begin{aligned}
a(z)\approx a'(1)(z-1)+\frac{1}{2}a''(1)(z-1)^2+\cdots,
\end{aligned}
\end{equation}
\begin{equation}\label{17}
 \begin{aligned}
\psi(z)\approx  \psi(1)+\psi'(1)(z-1)+\frac{1}{2}\psi''(1)(z-1)^2+\cdots.
\end{aligned}
\end{equation}
Next we intend to solve for $a''(1)$ and $\psi''(1)$ by series
expansion of the equations of motion. Substituting
$g(z)=\frac{{r_H}^2}{z^2} \left(1-z^4\right)$ into the EOM, we
obtain
 \begin{equation}\label{20}
 \begin{aligned}
 a''(1)=-\frac{-144{r_H}^2 \gamma +2 \mathcal{K} k^2+{r_H}^2\psi '(1)^2-2{r_H}^2}{2 \left(-24{r_H}^2 \gamma +\mathcal{K} k^2+{r_H}^2\right)}a'(1),
 \end{aligned}
\end{equation}

 \begin{equation}\label{21}
 \begin{aligned}
 \psi'(1)=\frac{3}{4}\psi(1),
 \end{aligned}
\end{equation}
and
 \begin{equation}\label{22}
 \begin{aligned}
\psi''(1)=-\frac{21}{16}\psi(1)-\frac{a'(1)^2\psi(1)}{32{r_H}^2}+\frac{9}{8}\psi'(1).
 \end{aligned}
\end{equation}
Combining (\ref{21}) and (\ref{22}), we have
 \begin{equation}\label{23}
 \begin{aligned}
\psi''(1)=-\frac{15}{32}\psi(1)-\frac{{a'(1)^2}\psi(1)}{32{r_H}^2}
 \end{aligned}.
\end{equation}
Finally, inserting (\ref{20}), (\ref{21}) and (\ref{23}) into
(\ref{16}) and (\ref{17}), we obtain the series expansion of the
fields as
 \begin{equation}\label{24}
 \begin{aligned}
a(z)\approx a'(1)(z-1)-\frac{-144{r_H}^2 \gamma +2 \mathcal{K} k^2+{r_H}^2\psi '(1)^2-2{r_H}^2}{2 \left(-24{r_H}^2 \gamma +\mathcal{K} k^2+{r_H}^2\right)}a'(1)(z-1)^2+\cdots,
 \end{aligned}
\end{equation}
and
 \begin{equation}\label{25}
 \begin{aligned}
\psi(z)\approx  \psi(1)+\frac{3}{4}\psi(1)(z-1)-\frac{1}{2}\left(\frac{15}{32}\psi(1)+\frac{{a'(1)^2}\psi(1)}{32{r_H}^2}\right)(z-1)^2+\cdots.
 \end{aligned}
\end{equation}

\subsubsection{Matching at $z=\frac{1}{2}$}
Matching (\ref{15}) with (\ref{24}) and (\ref{25}) at $z=\frac{1}{2}$, we obtain
 \begin{equation}\label{26}
 \begin{aligned}
\psi(1)^2=-\frac{2 \left(a'(1) \left(-120 {r_H}^2\gamma +3 \mathcal{K} k^2+{r_H}^2\right)+2 q \left(-24{r_H}^2 \gamma +\mathcal{K} k^2+{r_H}^2\right)\right)}{{r_H}^2 a'(1)},
 \end{aligned}
\end{equation}
and
 \begin{equation}\label{27}
 \begin{aligned}
a'(1)=-\frac{r_H \sqrt{145 \lambda_+ -126}}{\sqrt{\lambda_+ +2}}
 \end{aligned}.
\end{equation}
 Furthermore, inserting (\ref{27}) into (\ref{26}), we have
  \begin{equation}\label{28}
 \begin{aligned}
\psi(1)^2=240 \gamma -\frac{6 \mathcal{K} k^2}{{r_H}^2}-2 +\frac{4 \sqrt{\lambda +2} q \left(\mathcal{K} k^2+(1-24 \gamma ) {r_H}^2\right)}{\sqrt{145 \lambda_+ -126} {r_H^3}}.
 \end{aligned}
\end{equation}
After replacing $q$ by $\frac{\rho}{{r_H}^2}$ and
$r_H$ by $\pi{L^2}T$ respectively, we set $\psi(1)^2=0$,
then the critical temperature for condensation can be estimated by
finding the root of the following polynomial equation
\begin{equation}\label{29}
 \begin{aligned}
T^5-\frac{3 \mathcal{K} k^2 T^3}{\pi ^2 b}+\frac{2 (1-24 \gamma ) \sqrt{\lambda_+ +2} \rho }{\pi ^3 b c}T^2+\frac{2 \mathcal{K} k^2 \sqrt{\lambda_+  +2} \rho }{\pi ^5 b c}=0,
 \end{aligned}
\end{equation}

where $b=120 \gamma -1$ and $c=\sqrt{145 \lambda_+  -126}$. For
instance, if we set $\gamma=0, \mathcal{K}=\frac{1}{7}$ while
change $k/\rho^{1/3}=0, 1, 2$, we obtain the values of the
critical temperature $T/\rho^{1/3}$ as 0.20170, 0.17011, 0.14979,
respectively, implying that the condensation becomes harder with
the increase of momentum relaxation $k$. Moreover, if we fix for
instance $\rho=1, \mathcal{K}=\frac{1}{7}, k=1$ but change
$\gamma=-\frac{1}{96}, 0, \frac{1}{124}$, the corresponding
values for the critical temperature $T/\rho^{1/3}$ are
0.15534, 0.17011, 0.20854, implying that the increase of the Weyl
parameter makes the condensation easier. In next part we will see
soon that this trend can be justified by explicitly solving
the equations of motion with numerical method.

\subsection{Numerical part}

In this part we will explicitly solve the condensation equations
with numerical method and demonstrate the parameter dependence of
the critical temperature, then we will numerically solve the
perturbation equations to compute the optical conductivity along
x-direction below the critical temperature. In the remainder
of this paper, we will take the charge density as the unit such
that all the dimensionless quantities will be denoted with a
tilde, namely as $\tilde{O}$. Thus $\tilde{T}$ is the
dimensionless temperature $T/\rho^{1/3}$ while $\tilde{T_c}$ is
$T/\rho_c^{1/3}$ and so forth.
\begin{itemize}
\item Condensation of the scalar field. First we numerically
determine the critical temperature by solving EMO (\ref{11}) and
(\ref{10}), at which the scalar hair starts to condensate, leading
to nontrivial solutions. We solve these equations with the use of
the spectral method. As a result, the parameter dependence of the
critical temperature is shown in FIG.\ref{figureb1}. From the left
plot we learn that the critical temperature becomes lower with the
increase of $\tilde{k}$, but becomes higher with the increase
of $\gamma$, which is consistent with FIG.\ref{figure2}.
 These numerical results verify our analytical
approximation through the matching method in previous subsection.
In the presence of axions, as $\tilde{k}$ becomes larger, the
momentum dissipation becomes stronger such that the condensation
becomes harder, which is similar to the effects of the Q-lattices
as demonstrated in \cite{Ling:2014laa} but in contrast to the
scalar lattices as found in \cite{Horowitz:2013jaa}. Our result
here is also different from the previous superconductor models
with axions where the critical temperature may have non-monotonic
relation with $\tilde{k}$, for instance in
\cite{Andrade:2014xca,Kim:2015dna,Baggioli:2015zoa}. This is
simply because only the probe limit is taken into account in our
paper. On the other hand, when the Weyl correction is taken into
account, as $\gamma$ becomes larger, the generation of
electron-like excitations exceeds that of vortex-like
excitations
 such that the condensation of
s-wave becomes easier. What we have found for changing
$\gamma$ is consistent with the previous results as discussed in
\cite{Wu:2010vr,WitczakKrempa:2012gn}. However, it is worthwhile
to point out that in p-wave model the presence of Weyl term may
make the phase transition harder \cite{Zhao:2012kp}.

\begin{figure}[htbp]
  \centering
  \includegraphics[height=4.5cm]{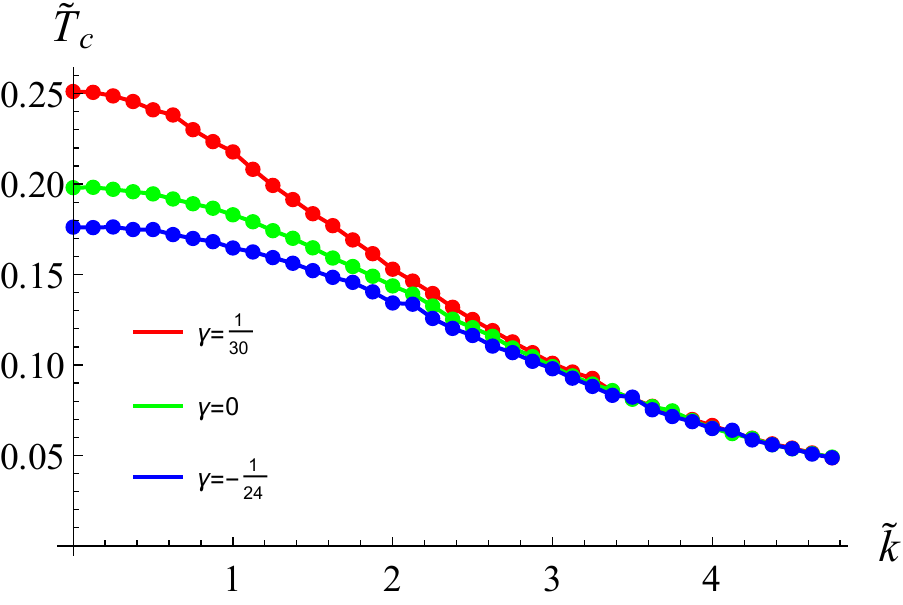}\qquad
    \includegraphics[height=4.5cm]{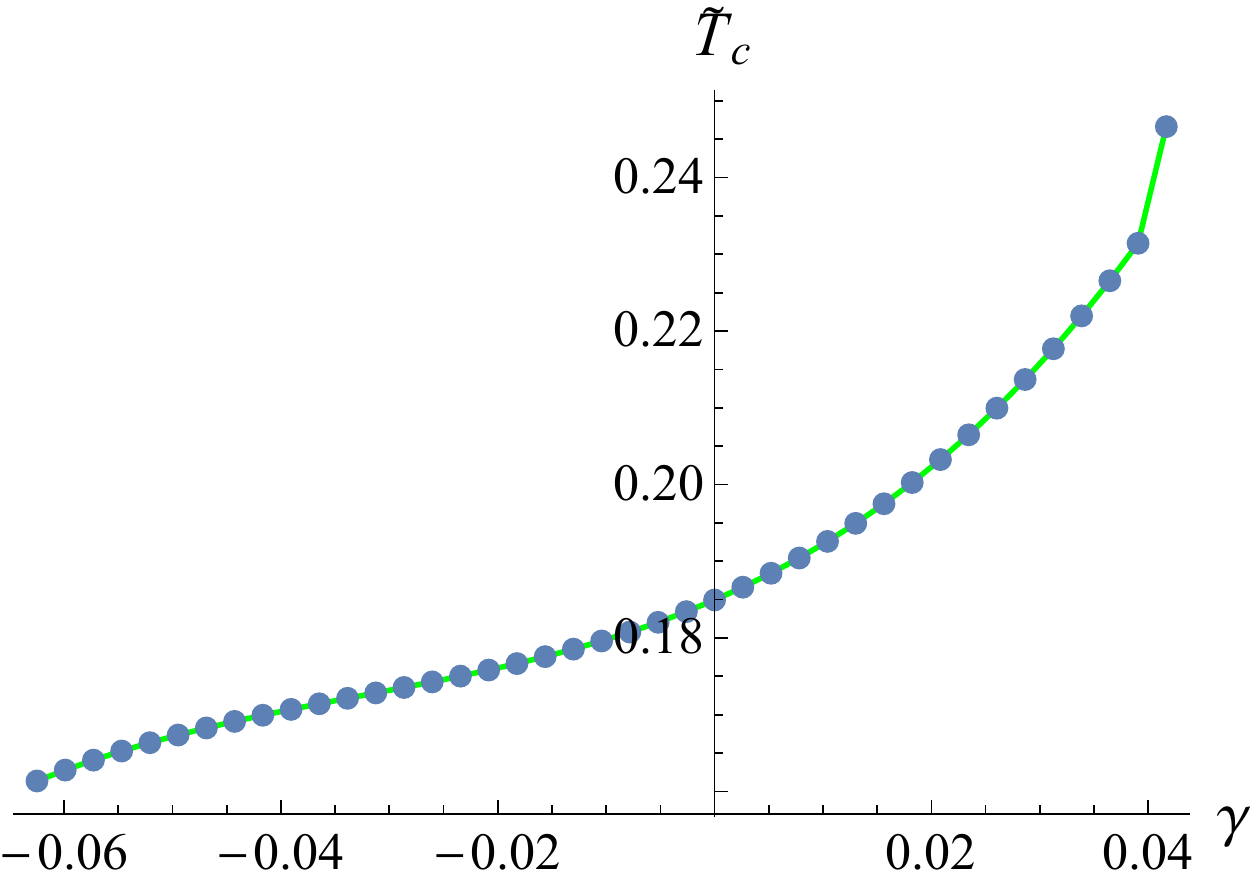}
 \caption{\label{figureb1}
The variation of the critical temperature with the system
parameters. The left plot is $\tilde{T_c}$ versus
$\tilde{k}$ for various $\gamma$, while the right plot is
$\tilde{T_c}$ versus $\gamma$ for $\tilde{k}=1$.}
\end{figure}

\begin{figure}[!h]
\centering
\includegraphics[height=4.5cm]{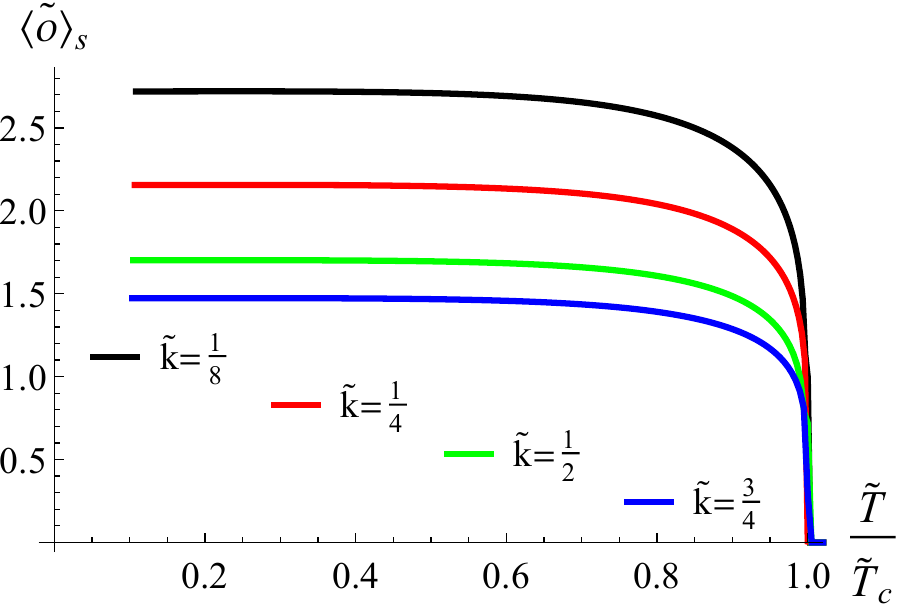}\qquad
  \caption{\label{figure2}
The condensation of the scalar field below the critical
temperature for various $\tilde{k}$ with $\gamma=0$}.
\end{figure}

\item The frequency behavior of the optical conductivity and
energy gap. Now we consider the transport behavior of the dual
system by linear response theory. We first derive the linear
perturbation equation for $A_x(z,t)=A_x(z)e^{i\omega t}$,
which reads as
\begin{equation}\label{f33}
\begin{aligned}
A_x''(z)+\left(\frac{3}{z}+\frac{g'(z)}{g(z)}+\frac{2 \left(\gamma
z^5 g^{(3)}(z)+6 \gamma  z^4 g''(z)+6 \gamma  z^3
g'(z)+3\right)}{z h(z)}\right)A_x'(z)\\+\left(\frac{6 \psi
(z)^2}{z^4 g(z) h(z)}+\frac{\omega ^2}{z^4 g(z)^2}\right)A_x(z)=0,
\end{aligned}
\end{equation}
where $h(z)=2 \gamma  z^4 g''(z)+8 \gamma  z^3 g'(z)+4 \gamma  z^2
g(z)-3 \mathcal{K} k^2 z^2-3$. Next we turn on the external
electric field along x-direction by imposing appropriate boundary
condition at $z=0$. As usual, the ingoing boundary condition is
imposed on the horizon. We demonstrate the frequency behavior of
the conductivity in Fig.\ref{figured1} and FIG.\ref{figure4} for
different values of parameters. First of all, we notice that the
imaginary part of the conductivity is divergent in zero frequency
limit, indicating a superconducting phase for the dual system.
Secondly, we observe that the energy gap shifts with the change of
the parameters in our model. Specifically, the energy gap becomes
small with the increase of $\tilde{k}$, as illustrated in
FIG.\ref{figured1}. By locating the minimal value of the imaginary
part of the conductivity,

 we find that the energy gap $\tilde{\omega}/\tilde{T}_c$
is 8.26, 8.13, 7.88, 7.63 corresponding to $\tilde{k}=\frac{1}{8},
\frac{1}{4}, \frac{1}{2}, \frac{3}{4}$, respectively. On the other
hand, the energy gap decreases with $\gamma$, as illustrated in
FIG.\ref{figure4}, where the value of energy gap runs from $7.51,
7.26, 7.00$ to $6.26$, corresponding to $\gamma=-\frac{1}{96}, 0,
\frac{1}{96}, \frac{1}{32}$, respectively.

What we have observed
above are consistent with the results in \cite{Wu:2010vr}, but
have an opposite tendency in comparison with the results in
Gauss-Bonnet gravity and quasi-topological gravity in which the
energy gap becomes larger than $8$ with the increase of system
parameters\cite{Pan:2009xa,Kuang:2010jc}.

\item A modified Homes' law. In the presence of the axions , the
dual system is a two-fluid system below the critical temperature,
composed of the normal fluid and the superfluid. The density of
superfluid can be evaluated by $Im(\tilde{\sigma}) \sim 2\pi
\tilde{\rho}_s/\tilde{\omega}$ with extremely low temperature. As
a consequence, we may investigate the relation between this
quantity and $\tilde{\sigma} _{DC}(\tilde{T_c}) \tilde{T_c}$ by
changing system parameters. Some examples have been
illustrated by changing $\tilde{k}$ in FIG.\ref{figuree1}, with
$\gamma=-\frac{1}{24}, 0, \frac{1}{30}$, respectively. From right
to left, $\tilde k$ runs from 0 to 5 for each value of
$\gamma$. We remark that here both $\tilde{\rho}_s$ and
$\tilde{\sigma} _{DC}(\tilde{T_c}) \tilde{T_c}$ are scaleless
quantities. Interestingly enough, a manifest linear relation has
been observed for these two quantities, signalizing a Homes' law
for this model. However, our data fitting tells us that the
intercept at the vertical axis might not be zero if extending the
straight line to $\tilde{\sigma} _{DC}(\tilde{T_c})
\tilde{T_c}=0$. Instead, a modified version of Homes' law is the
best fitting, which is $\tilde{\rho}_s =C \tilde{\sigma}
_{DC}(\tilde{T_c}) \tilde{T_c} + a$. In this figure for different
$\gamma$, we have $\tilde{\rho} _s =0.90 \tilde{\sigma}
_{DC}(\tilde{T_c}) \tilde{T_c} +1.26$, $\tilde{\rho} _s =0.94
\tilde{\sigma} _{DC}(\tilde{T_c}) \tilde{T_c} +1.23$,
$\tilde{\rho} _s =0.96 \tilde{\sigma} _{DC}(\tilde{T_c})
\tilde{T_c} +1.20$, respectively.
 It is worthwhile to point out that we are not able to evaluate
$\tilde{\rho}_s$ at absolute zero temperature numerically. Instead
we obtain its value at $\tilde{T}=0.1\tilde{T_c}$. This
identification should be fine since
$\tilde{\rho}_s\thicksim\tilde{\langle\mathcal {O}\rangle}$
\cite{Hartnoll:2008vx,Horowitz:2009ij} and from Fig.\ref{figure2}
we know $\tilde{\rho}_s$ is almost saturated to a constant below
the critical temperature. In comparison with the results presented
in \cite{Erdmenger:2015qqa}, we find the dimensionless constant
$C$ in our model is about one and much smaller than those observed
in laboratory. We propose that different values for $C$ could be
obtained by introducing more general coupling terms or potentials
of axions into this simple model. Numerically we are not able to
touch lower temperature region for more data.
But from this figure we also notice that the data have tendency to
go down quickly and then would deviate from a linear relation in
low temperature region, in particular, for those dots in red. We
also conjecture that perhaps a more cautious consideration, for
instance with full backreactions to the background, is needed in
this low temperature limit.

Finally, we present our preliminary understanding on
the non-vanishing constant $a$ as follows. Assume that the
density of superfluid is a function of temperature as well as
$\tilde{k}$, i.e., $\tilde{\rho_s}(\tilde{T},\tilde{k})$. $a\neq
0$ implies that $\tilde{\rho_s}$ would be $\tilde{\rho_s}=a$ for
some $\tilde{k}$ at $\tilde{T}=0$ when $\tilde{T_c}\rightarrow0$
such that $\tilde{\rho_s}$ as a function of $\tilde{T}$ would be
discontinuous at $\tilde{T}=0$, i.e.,
\begin{eqnarray}
\tilde{\rho_s}\xlongequal{\tilde{T}_c\rightarrow0}
\left\{
\begin{array}{c}
a,    \quad \tilde{T}=0\\
0,    \quad \tilde{T}\neq0
\end{array}
\right.
.
\end{eqnarray}
Interestingly enough, this implies that
$\tilde{\rho}_s\thicksim\tilde{\langle\mathcal {O}\rangle}$ \cite{Hartnoll:2008vx,Horowitz:2009ij}
would be discontinuous at $\tilde{T}=0$, which signalize the
derivative of free energy $\tilde{\langle\mathcal
{O}\rangle}\thicksim\frac{\delta S}{\delta\phi_0}\mid_{\phi_0=0}$
is discontinuous at $\tilde{T}=\tilde{T_c}=0$. It implies that
superconducting transition would be the first order phase
transition at $\tilde{T_c}=0$ rather than the second order one
(But it is the second order phase transition whenever
$\tilde{T_c}\neq0$). Furthermore, the non-vanishing signature of
the density of superfluid for $a\neq0$ might reflect the property
of quantum critical phenomenon. In fact, one can fix
$\tilde{T}=0$, then $\tilde{\rho}_s$ would simply be a function of
$\tilde{k}$ ranging from $\tilde{k}=-\infty$ to
$\tilde{k}=\infty$. Then $\tilde{\rho}_s(\tilde{T}=0,\tilde{k})$
would be discontinuous at some point $\tilde{k_c}$ corresponding
to $\tilde{T_c}=0$. Anyway, we intend to stress that
the non-vanishing $a$ would be an artifact of the probe limit.
Since the backreaction is very complicated in particular in the
presence of the Weyl term, we would like to leave this issue for
future investigation.

\end{itemize}

\begin{figure}[htbp]
  \centering
  \includegraphics[height=4.5cm]{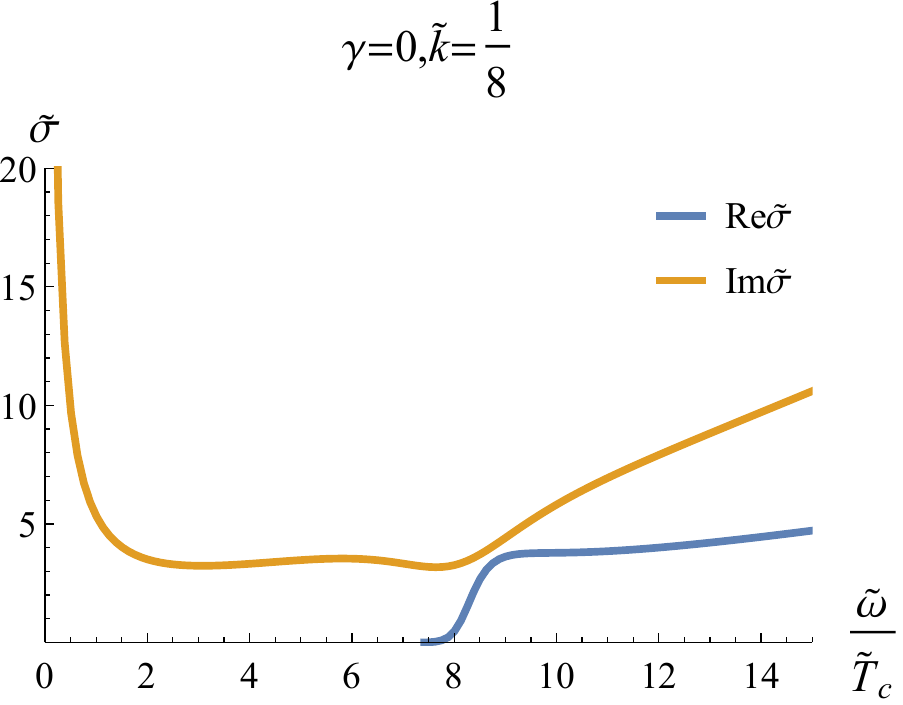}\qquad
  \includegraphics[height=4.5cm]{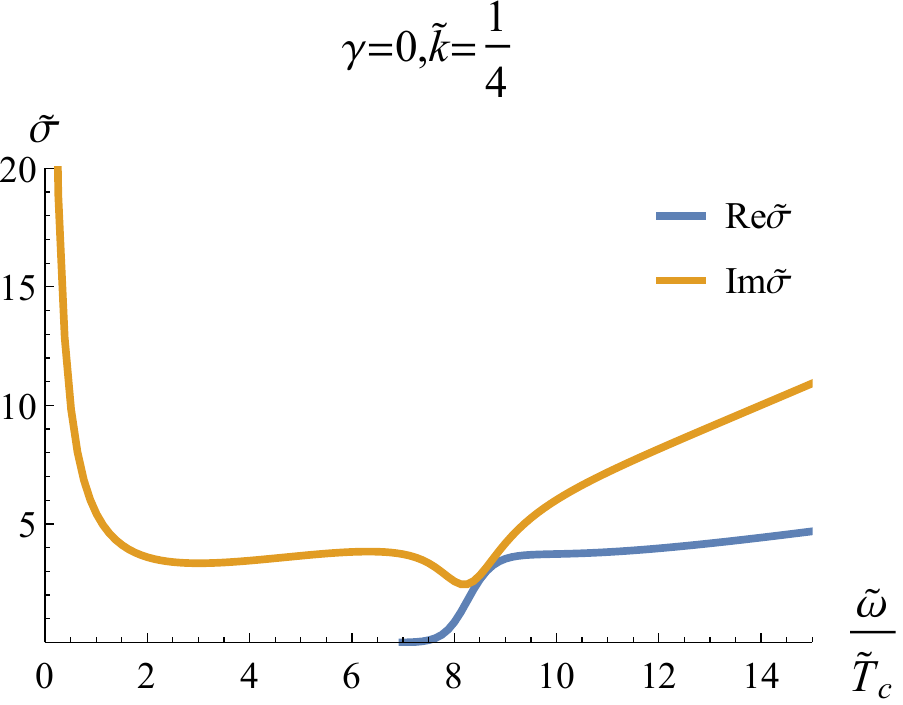}\qquad
  \includegraphics[height=4.5cm]{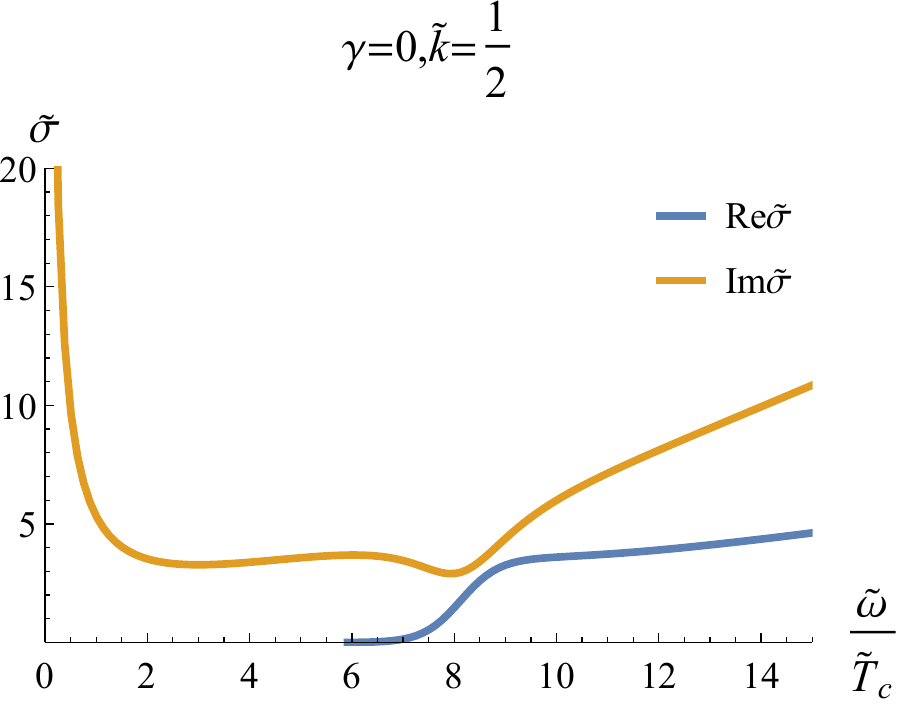}\qquad
  \includegraphics[height=4.5cm]{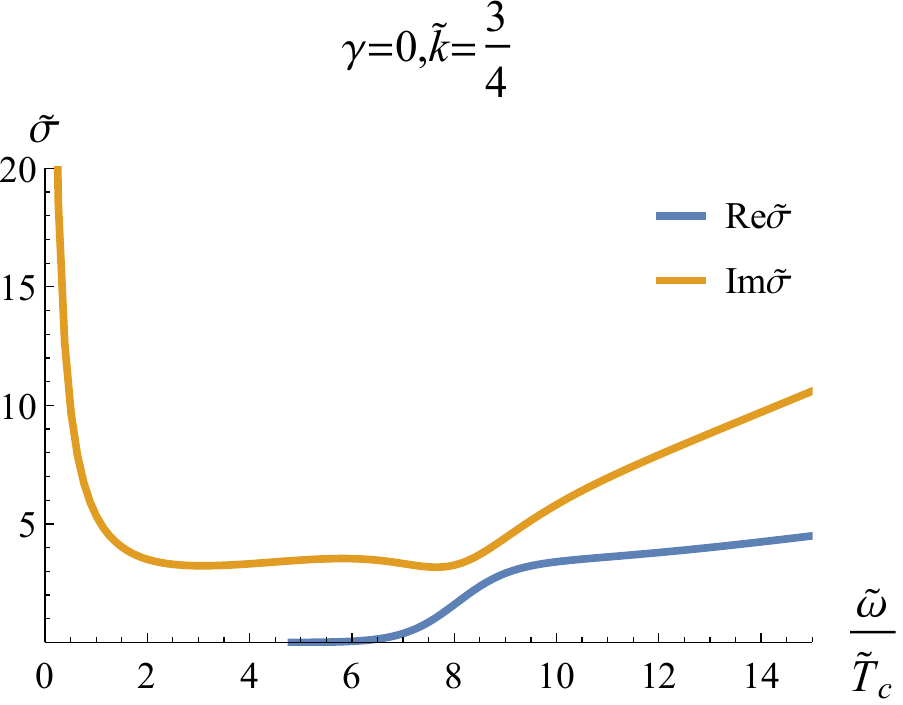}
  \caption{\label{figured1}
The frequency behavior of the optical conductivity for various
values of $\tilde{k}$, with $\gamma=0$ and $\tilde{T}/\tilde{T_c}=0.1$
fixed.}
\end{figure}
\begin{figure}[htbp]
  \centering
  \includegraphics[height=4.5cm]{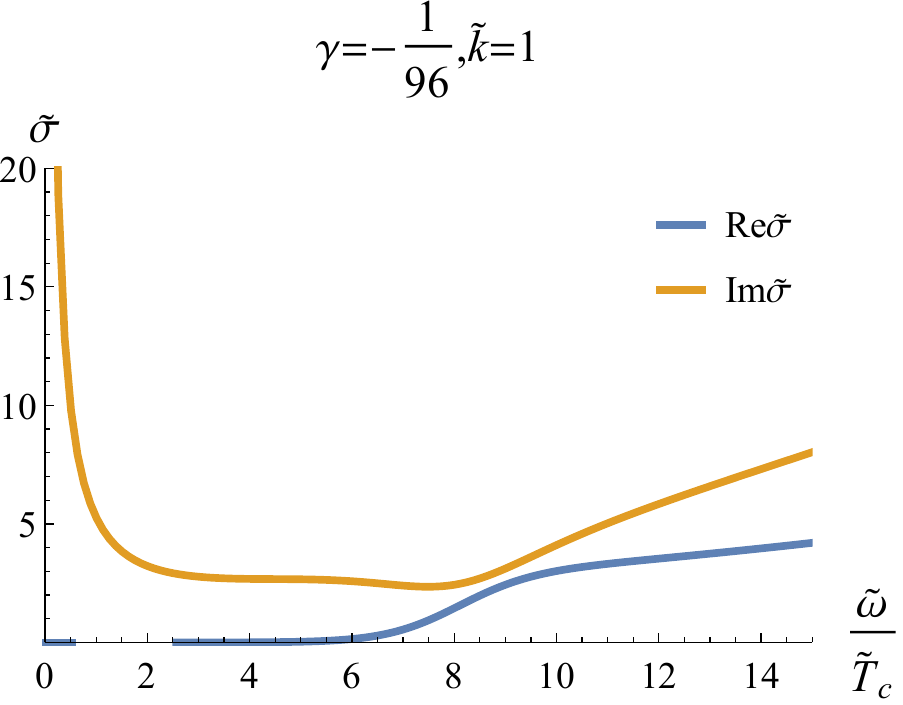}\qquad
    \includegraphics[height=4.5cm]{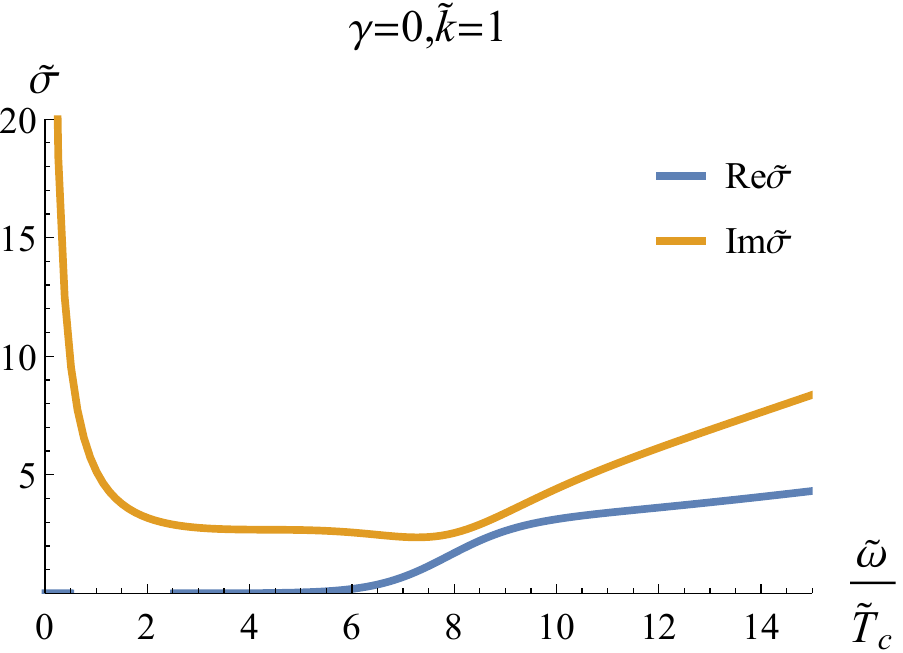}\qquad
  \includegraphics[height=4.5cm]{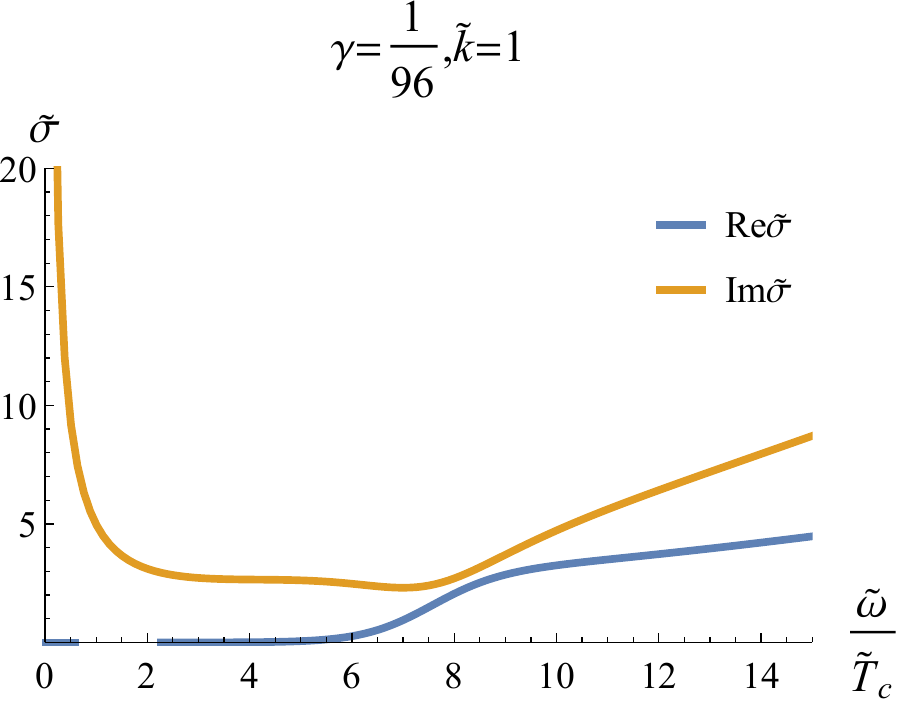}\qquad
  \includegraphics[height=4.5cm]{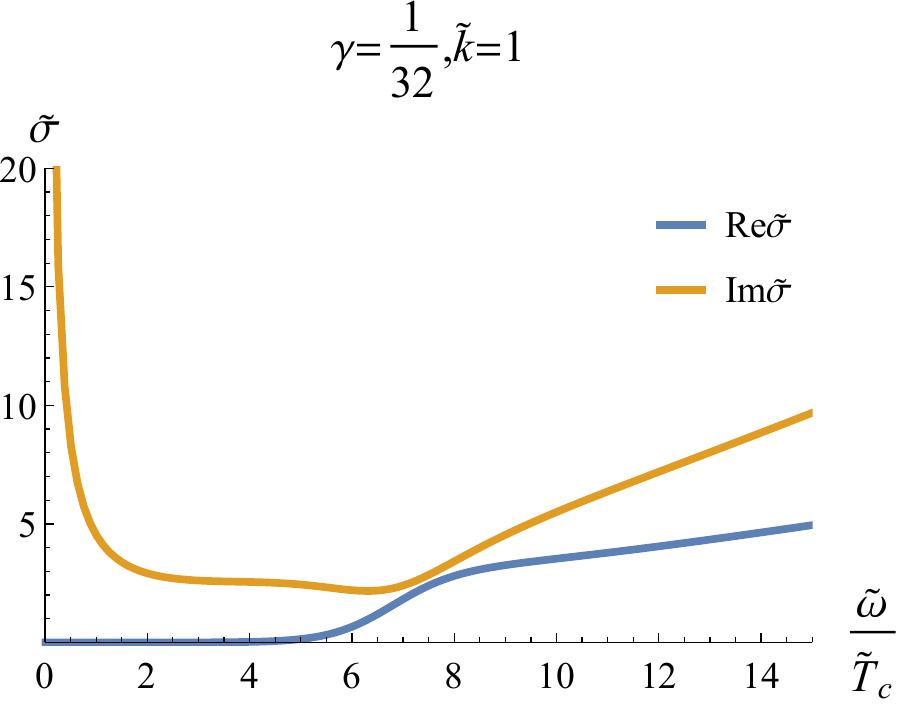}
  \caption{\label{figure4}
The frequency behavior of the optical conductivity for various
values of $\gamma$, with $\tilde{k}=1$ and $\tilde{T}/\tilde{T_c}=0.1$
fixed.}
\end{figure}

\begin{figure}[htbp]
  \centering
  \includegraphics[height=6.5cm]{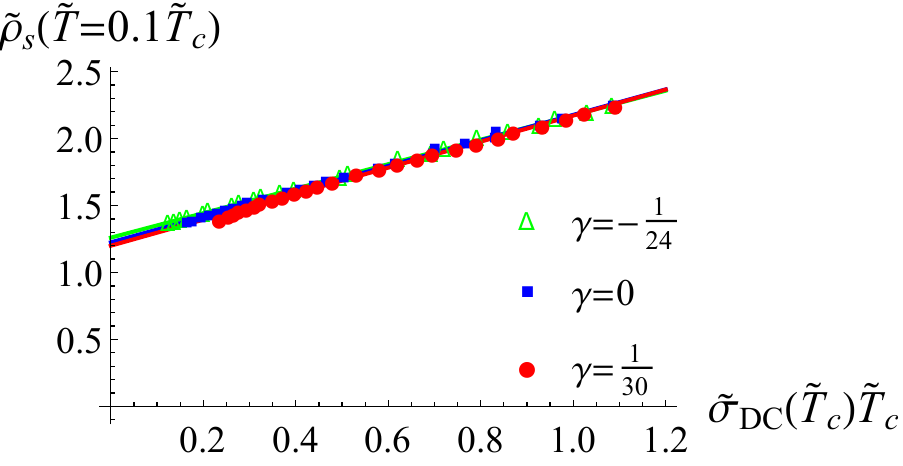}
  \caption{\label{figuree1}
The plot is for $\tilde{\rho} _s (\tilde{T}=0.1\tilde{T_c})$ versus $\tilde{\sigma}
_{DC}(\tilde{T_c})\tilde{T_c}$.
 A linear relation is observed with a fixed $\gamma$ and
our fitting gives rise to a modified Home's law $\tilde{\rho} _s=C
\tilde{\sigma} _{DC}(\tilde{T_c})\tilde{T_c}+a$.}
\end{figure}

\section{Discussion}\label{section5}

In this paper we have constructed a holographic model with Weyl
corrections in five dimensional spacetime. Momentum relaxation is
introduced by the coupling term between the axions and the Maxwell
field. For the normal state of the conductivity we find the
portion of incoherence is suppressed with the increase of the
strength of axions, which is in contrast to the previous
holographic models with momentum relaxation induced by axions. For
the superconducting phase we have found that the critical
temperature decreases with $\tilde{k}$, indicating that the
condensation becomes harder in the presence of the axions, while
it increases with Weyl parameter. More importantly we find the
density of superfluid at zero temperature has a linear relation
with the quantity $\tilde{\sigma} _{DC}(\tilde{T_c})\tilde{T_c}$
which can be described by a modified formula of Homes' law.
Moreover, constant $a\neq0$ implies superconducting
transition would be the first order phase transition at
$\tilde{T_c}=0$ rather than the second order one and
would reflect the property of quantum critical
phenomenon.

Some crucial issues are worth for further investigation. First of
all, due to the presence of the Weyl correction, we have only
investigated the conductivity of the dual system in the probe
limit. It is very interesting to take the backreactions of matter
fields to the background into account, which leads to higher order
differential equations of motion\cite{Myers:2010pk,Ling:2016dck}.
In this situation we expect that more abundant phenomena could be
observed for the transport behavior of the dual system. In
particular, some quantum critical phenomenon such as
metal-insulator transition can be implemented and its relation
with the holographic entanglement entropy could be investigated as
explored in \cite{Ling:2016dck}. Secondly, we have only considered
a special coupling term between the axion fields and Maxwell
field, it is quite intriguing to construct more realistic models
which could be described by the Homes' law with a constant
compatible with the experimental data. Finally, the incoherent
part of the conductivity can approximately be described by a
constant $\tilde{\sigma}_Q$  only in low frequency region under the
condition that the coherent contribution is dominant. It is very
worthy of investigating its frequency dependent behavior in a
generic situation.

\section*{Acknowledgements}

We are very grateful to Wei-jia Li, Peng Liu, Zhuoyu Xian, Jianpin
Wu and Zhenhua Zhou for helpful discussion. We also thank W.
Witczak-Krempa for his very valuable comments on the previous
version of the manuscript. This work is supported by the Natural
Science Foundation of China under Grant Nos.11275208 and 11575195,
and by the grant (No. 14DZ2260700) from the Opening Project of
Shanghai Key Laboratory of High Temperature Superconductors. Y.L.
also acknowledges the support from Jiangxi young scientists
(JingGang Star) program and 555 talent project of Jiangxi
Province.

\section*{Appendix}\label{section6}
In this appendix we present the detailed derivation from
(\ref{a40}) to (\ref{a41}). Without loss of generality, we may
start from the perturbation equation (\ref{f33}) with
$\gamma=k=0$, then we have

\begin{equation}\label{a54}
\begin{aligned}
A_x''(z)+\frac{A_x'(z) \left(z^4 g(z) g'(z)+z^3 g(z)^2\right)}{z^4 g(z)^2}+\frac{A_x(z) \left( \omega ^2-2 g(z) \psi (z)^2\right)}{ z^4 g(z)^2}=0.
\end{aligned}
\end{equation}
Since the bulk geometry is asymptotically AdS,  the field
$A_x(z)$ has the following expansion near the boundary $z=0$.
\begin{equation}\label{a55}
\begin{aligned}
 A_x(z)=A_x^0+A_x^2 z^2-\frac{1}{2} A_x^0 \omega ^2 z^2 \log (\Lambda  z)+\cdots,
\end{aligned}
\end{equation}
For convenience, we set $A_x^0=1$, then obtain
\begin{equation}\label{a56}
\begin{aligned}
 A_x(z)=1+A_x^2 z^2-\frac{1}{2}  \omega ^2 z^2 \log (\Lambda  z)+\cdots,
\end{aligned}
\end{equation}
Taking the ingoing boundary condition at $z=1$ into account,
one usually defines ${A_x}(z)=(1-z)^{\frac{-i\omega}{4}}{b_x}(z)$. Then inserting it into (\ref{a54}), we have the
equation for ${b_x}(z)$ as following
\begin{equation}\label{a58}
\begin{aligned}
{b_x}''(z)+c(z){b_x}'(z)+d(z){b_x}(z)=0,
\end{aligned}
\end{equation}
where
 \begin{equation}\label{a59}
\begin{aligned}
c(z)=\frac{2 (z-1) z g'(z)+g(z) (2 (z-1)-i \omega  z)}{2 (z-1) z g(z)}
\end{aligned}
\end{equation}
and
 \begin{equation}\label{a60}
\begin{aligned}
d(z)=&\frac{16 \omega ^2 (z-1)^2+\omega  z^3 g(z)^2 (-\omega  z+4 i )}{16  (z-1)^2 z^4 g(z)^2}
\\&-\frac{4 (z-1) g(z) \left(8 (z-1) \psi (z)^2+i \omega  z^4 g'(z)\right)}{16 (z-1)^2 z^4 g(z)^2}.
\end{aligned}
\end{equation}
Correspondingly, we find the asymptotic expansion of ${b_x}(z)$
near the boundary to be
 \begin{equation}\label{a61}
\begin{aligned}
{b_x}(z)=1-\frac{i \omega  z}{4 }+z^2 \left(A_x^2-\frac{\omega
^2}{32}-\frac{i \omega }{8}\right)-\frac{1}{2} \omega ^2 z^2 \log
(\Lambda z)+\cdots.
\end{aligned}
\end{equation}
Due to the presence of the logarithmic term which is
non-analytical, it is not quite efficient to numerically solve the
ODE with Chebyshev polynomials. In particular, it is not quite
convenient to extract the coefficients in (\ref{a56}) to calculate
the Green's function for conductivity. Thus, to improve the
efficiency in numerics, we find it is convenient to introduce a
new variable $a_x(z)$ by setting $b_x(z)=-\frac{1}{2} \omega
^2 z^2 \log (\Lambda  z)+a_x(z)$, then we find the asymptotic
behavior of $a_x(z)$ is
\begin{equation}\label{a66}
\begin{aligned}
{a_x}(z)=1-\frac{i \omega  z}{4}+z^2 \left(A_x^2-\frac{\omega ^2}{32}-\frac{i \omega }{8}\right)+\cdots,
\end{aligned}
\end{equation}
then we get
\begin{equation}\label{a67}
\begin{aligned}
 \frac{2 A_x^2}{A_x^0}=a_x''(0)+\frac{\omega ^2}{16}+\frac{i \omega
 }{4}.
\end{aligned}
\end{equation}
Substituting (\ref{a67}) into (\ref{a40}), the formula of
conductivity can be expressed in terms of the derivatives of
the new variable $a_x(z)$ as
\begin{equation}\label{a68}
\begin{aligned}
\sigma (\omega )=-\frac{i \left(16  a_x''(0)+16 \omega ^2 \log \left(\frac{\Lambda }{\nu }\right)-7  \omega ^2+4 i \omega \right)}{16 \omega }.
\end{aligned}
\end{equation}
We would like to remark that $a_x(0)=1$, otherwise $a_x(z)$
would also have nonanalytic terms between the source term and
the response term. Similarly, if the expansion of fields have
other nonanalytic terms such as $z^k$ with {\it non-integer} $k$
between source and response, we can get rid of these nonanalytic
terms in a parallel way. Finally, this method can be
straightforwardly generalized to solve equation group with
multiple variables.

\end{document}